\documentclass{aa}  

\usepackage{graphicx}
\usepackage{txfonts}
\usepackage{textcomp}
\usepackage{natbib}
\bibpunct{(}{)}{;}{a}{}{,}

\begin{document}

   \title{Transmission spectroscopy of HAT-P-32b with the LBT\thanks{Based on observations made with the Large Binocular Telescope (LBT) and the STELLA robotic telescopes. The LBT is an international collaboration among institutions in the United States, Italy and Germany. LBT Corporation partners are: LBT Beteiligungsgesellschaft, Germany, representing the Leibniz Institute for Astrophysics Potsdam (AIP), the Max-Planck Society,  and Heidelberg University; The University of Arizona on behalf of the Arizona Board of Regents; Istituto Nazionale di Astrofisica, Italy;  The Ohio State University, and The Research Corporation, on behalf of The University of Notre Dame, University of Minnesota and University of Virginia. STELLA is an AIP facility jointly operated by AIP and Instituto de Astrof\'{i}sica de Canarias (IAC).}: confirmation of clouds/hazes in the planetary atmosphere}
   \titlerunning{Transmission spectroscopy of HAT-P-32b}


   \author{M. Mallonn and K. G. Strassmeier}

   \authorrunning{M. Mallonn \& K. G. Strassmeier.}

   \institute{Leibniz-Institut f\"{u}r Astrophysik Potsdam, An der Sternwarte 16, D-14482 Potsdam, Germany\\
              \email{mmallonn@aip.de}
             }

   \date{Received 4 December 2015; accepted 29 March 2016 }

 
  \abstract
   {}
   {Spectroscopic observations of a transit event of an extrasolar planet offer the opportunity to study the composition of the planetary atmosphere. This can be done with comparably little telescope time using a low-resolution multi-object spectrograph at a large aperture telescope. We observed a transit of the inflated hot Jupiter HAT-P-32b with the Multi-Object Double Spectrograph
 at the Large Binocular Telescope  to characterize its atmosphere from 3300 to 10000~\AA{}.}
   {A time series of target and reference star spectra was binned in two broad-band wavelength channels, from which differential transit light curves were constructed. These broad-band light curves were used to confirm previous transit parameter determinations. To derive the planetary transmission spectrum with a resolution of $\mathrm{R} \sim 60$, we created a chromatic set of 62 narrow-band light curves. The spectrum was corrected for the third light of a nearby M star. Additionally, we undertook a photometric monitoring campaign of the host star to correct for the influence of starspots.}
   {The transmission spectrum of HAT-P-32b shows no pressure-broadened absorption features from Na and K, which is interpreted by the presence of clouds or hazes in the planetary atmosphere. This result is in agreement with previous studies on the same planet. The presence of TiO in gas phase could be ruled out. We find a 2.8\,$\sigma$ indication of increased absorption in the line core of potassium (K\,I~7699~\AA{}). No narrow absorption features of Na and H$\alpha$ were detected. Furthermore, tentative indications were found for a slope of increasing opacity toward blue wavelengths from the near-IR to the near-UV with an amplitude of two scale heights. If confirmed by follow-up observations, it can be explained by aerosols either causing Mie scattering or causing Rayleigh scattering with an aerosol - gas scale height ratio below unity. The host star was found to be photometrically stable within the measurement precision.}
   {}


   \maketitle
%

\section{Introduction}
Transiting extrasolar planets bright enough for ground-based follow-up provide a wealth of information for their characterization. Photometric observations of multiple transit events provide the orbital period, the orbital inclination, and the planet--star radius ratio. High-precision stellar spectroscopy yields the stellar parameters, and in comparison with stellar evolutionary models, the true stellar mass and size. These pieces of information combined allow the true size of the planet to be derived, which in turn provides a planetary bulk density when combined with information from the radial velocity curve of the host star. 

The observation of a planetary transit at multiple wavelengths allows for the characterization of the atmosphere of the planet, which can be used to understand planet formation and evolution and, ultimately, to search
for biomarkers. During a transit, a fraction of the starlight shines through the atmosphere of the planet where atoms and molecules scatter and absorb it dependent on wavelength. This effect is in principle measurable as a wavelength dependent effective radius.

The first observational attempts of transmission spectroscopy of an extended gas envelope surrounding an extrasolar planet were made in the infrared shortly after such planets were detected \citep{Coustenis1997}. The first experiments and the detection of upper limits on planetary absorption at optical wavelengths followed soon thereafter \citep{Rauer2000,Bundy2000,Moutou2001}. The successful detection of the absorption of sodium was announced by \cite{Charbonneau2002}. Since then, optical transmission spectroscopy has found atomic signatures not only of sodium \citep{Snellen2008,Sing2008,Redfield2008,Wood2011,Sing2012,Nikolov2014,Murgas2014} and potassium \citep{Sing2011_XO2,Nikolov2015,Wilson2015,Sing2015}, but also of atomic hydrogen H$\alpha$ \citep{Jensen2012}, calcium \citep{Astudillo2013}, and molecular hydrogen \citep{Sing2008,LecavelierDesEtangs2008}. 

Several planets exhibit a Rayleigh slope of increased absorption at blue wavelengths \citep{Pont2013,Sing2013,Jordan2013,Sing2015,Nikolov2015}, others show a flat and featureless spectrum \citep{Gibson2013,GibsonH32,MallonnH12,MallonnH19,Lendl2015}. Generally, the current results show a rather diverse picture of  hot Jupiter atmospheres. For example, clouds seem to play an important role in the interpretation of the atmospheres of many objects \citep{Sing2016}, but multiple objects exist that are explained by cloud-free models \citep[e.g.,][]{Fischer2016}. We do not know yet about the physical origin of the difference between these groups. Additionally, in principle all measurements need to deal with instrument systematics, i.e., to some extent, the diversity of current results might also be influenced by the low signal-to-noise ratio of most attempts of exoplanet spectroscopy. We still need to verify many of the detections by repeated observations of the most favorable targets.

The best chance for a detection of an atmospheric signal is provided by close-in gas giants with inflated radii and a large transit depth. The hot temperatures in their atmospheres and the low surface gravity drive the atmospheric pressure-scale height up to 1000~km and more for the most suitable targets. Among them is the planet HAT-P-32b. This hot Jupiter was among the first exoplanet candidates of the HATNet survey \citep{Bakos2004} with discovery photometry dating back to 2004 \citep{Hartman2011}. However, the object proved difficult to confirm because of a high radial velocity (RV) jitter of about $\sim$\,80~ms$^{-1}$. HAT-P-32b has a mass of about 0.9~M$_{\rm{Jupiter}}$ and a radius of about 1.8~R$_{\rm{Jupiter}}$, making it one of the most bloated planets found to date. The host is an F-type main-sequence star of V\,=\,11.3~mag, with a mass of 1.16~M$_{\odot}$ and a radius of 1.22~R$_{\odot}$. Although many more high-S/N spectra were taken for the RV curve than for typical HATNet 
candidates, the planet's eccentricity $e$ could not be well constrained owing to the jitter, $e = 0.2^{+0.19}_{-0.13}$ \citep{Knutson2014}. \cite{Hartman2011} argued that the high RV jitter is probably not caused by brightness (temperature) inhomogeneities on the stellar surface. Instead, it might be caused by convective inhomogeneities that vary in time. The photometric detection of the secondary eclipse by \cite{Zhao2014} and its timing increased the likelihood of a circular orbit. The same authors found a featureless emission spectrum of the planet's dayside in agreement with a spectrum of a black body of $\sim$\,2000~K.

\cite{Knutson2014} confirmed a trend in the RV measurements of HAT-P-32 that indicated an additional companion at a distance of several AU and with a possible mass of up to 0.5~M$_{\odot}$. \cite{Adams2013} imaged a nearby star that is too distant to explain the RV trend. Hence, it might be that HAT-P-32 has two massive outer companions and we have no knowledge of the third-light contribution of the inner one. \cite{Seeliger2014} conducted an analysis of transit timing variations (TTV) to search for additional close-in planets in low-integer resonant periods to HAT-P-32b which, in part, might be responsible for the RV jitter. Their null-result ruled out nearby planets down to super-Earth masses, but was not sensitive to the two massive bodies farther out. 

A first transmission spectrum of HAT-P-32b at optical wavelengths was obtained by \cite{GibsonH32} (hereafter G13). They used two Gemini-North GMOS transit observations to rule out pressure-broadened absorption features of sodium and potassium. The achieved featureless spectrum could be best explained by a cloud coverage of the planetary atmosphere. Current attempts of ground-based transmission spectroscopy are at the limit of the instrument's capabilities and conclusions are vulnerable to systematics in the light curves. Therefore, we were interested whether the results of G13 could be reproduced with a new data set and an independent analysis. Additionally, we monitored the host star HAT-P-32 with STELLA/WiFSIP during two observing seasons to be sure that the transit parameters are not affected by starspots.

This paper is structured as follows. Section~2 describes the observations and the data reduction. Section~3 and~4 present the analysis and results of the monitoring campaign and the third light of the M dwarf companion, respectively. The analysis and results of transit light curves are given in Section~5, followed by the discussion in Section~6, and the conclusions in Section~7.


\section{Observations and data reduction}

\subsection{Spectroscopic transit observation}

\begin{figure}
   \centering
    \includegraphics[height=\hsize]{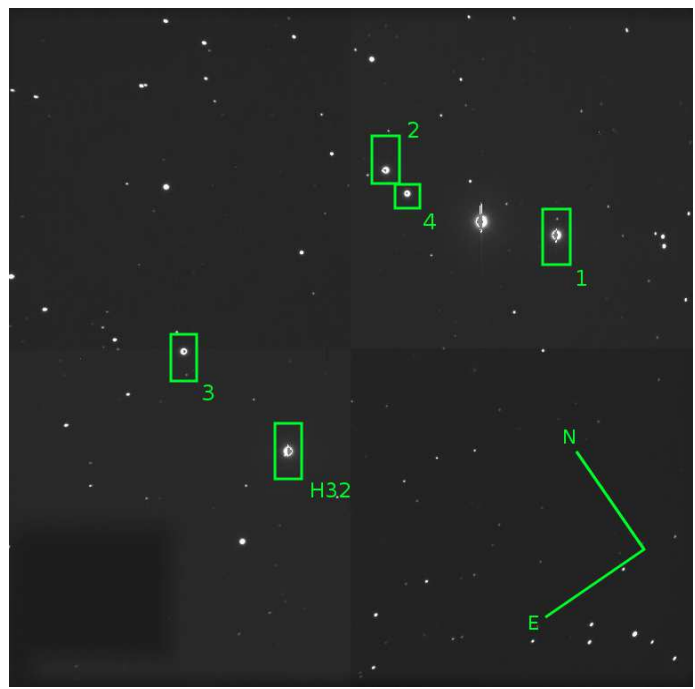}
      \caption{MODS acquisition image with 6\,$'\times$6\,$'$ field of view. A Sloan r' filter was used with all objects of interest saturated. The mask alignment was done by fainter stars not indicated here. The green boxes mark the approximate position of the slits in the mask, the dispersion direction is along the x-axis.}
         \label{ima_aqui_H32}
   \end{figure}

The transit observation was performed on 2012 November~13 with the Multi-Object Double Spectrograph \citep[MODS;][]{Pogge2010}, a low- to medium-resolution optical spectrograph/imager at the Large Binocular Telescope (LBT). MODS covers a field of view (FoV) of 6\,$'\times$6\,$'$. We employed the instrument in its dual-channel mode, i.e., the incoming light beam is split by a dichroic at a wavelength of 5650~\AA{} into separate red- and blue-optimized channels. A wavelength region from 3300 to 10000~\AA{} is covered in total. We used the multi-object spectroscopy (MOS) mode  where multiple stars are observed simultaneously through a user-designed slit mask. In this way we took spectra of the target and four comparison stars summarized in Table~\ref{H32_compstars}. Figure~\ref{ima_aqui_H32} shows the acquisition image of the field prior to the slit mask insertion. As dispersive element we chose the gratings G400L (blue) and G670L (red). Both channels are equipped with e2v 3K$\times$8K CCD detectors, 
differently coated for the blue and the red arm, which are read-out through eight amplifiers each. In November 2012 the telescope control software was not yet able to allow for binocular observations with MODS and so we used the binocular telescope in ``one-eye'' mode. For high-accuracy spectrophotometry we wanted to be sure of losing a negligible amount of flux at the edge of the slits. Owing to pointing and seeing variations this light loss would be variable and could potentially manifest as a significant noise source in the resulting light curves. Therefore, we designed 10\,$''$ wide slits for the MOS mask. The length of the slits was 30\,$''$ to allow for a proper sky correction.

\begin{figure}
   \centering
   \includegraphics[width=\hsize]{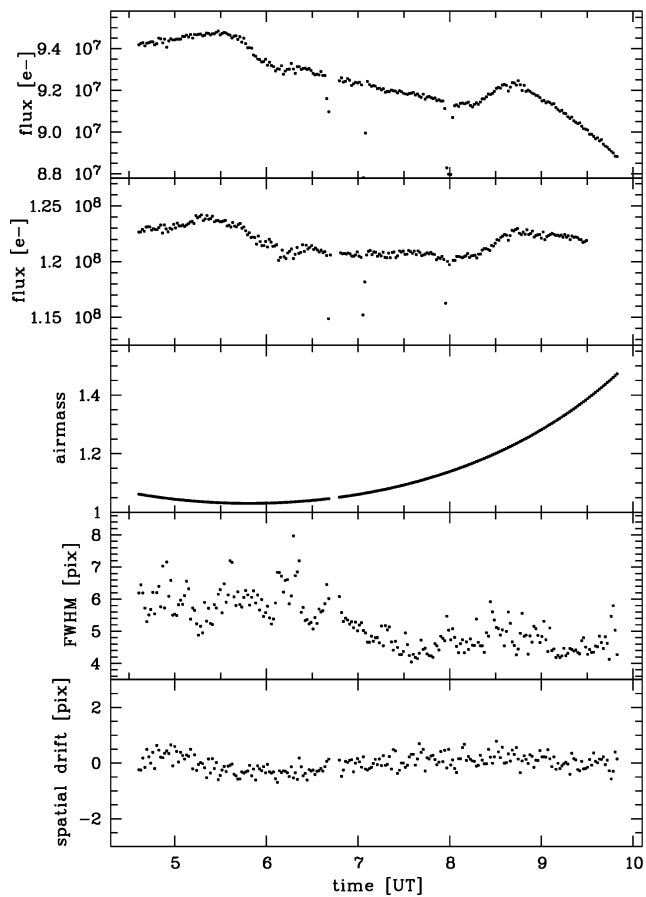}
      \caption{Evolution of observational parameters over the time series. From top to bottom: The count rate of HAT-P-32 of the blue light curve and the red light curve, the airmass, the full width at half maximum (FWHM) of the spectral profile in spatial direction, and the drift of the centroid of the spectral profile in spatial direction. }
         \label{plot_obs_cond}
   \end{figure}

The MODS observation started at 04:36~UT and lasted until 09:49~UT. The last exposures with the red arm could not be transferred to the data archive because of technical problems, hence the red time series ended at 09:29~UT. The transit took place between 05:43 and 08:49~UT. At 06:41~UT the observation paused owing to technical problems with the adaptive secondary mirror, but were continued after $\sim$\,7~min. All spectroscopic frames were exposed for 20~s, together with the overheads of 54 and 59~s resulting in a cadence of 74 and 79~s for the red and blue arm, respectively. A 2$\times$2~pixel binning was applied on-chip. The time series consisted of 238 blue frames and 240 red frames. The night was clear and offered photometric conditions; the seeing varied between 1.0\,$''$ and 1.5\,$''$. The airmass increased from 1.03 to 1.39 during the observation. The object drifted by less than two pixels in both spatial and dispersion direction over the length of the time series. 
The time dependence of these observational parameters are presented in Figure~\ref{plot_obs_cond} together with the raw flux of the target in the blue and the red arm. For each exposure the target spectra achieved a signal-to-noise of about 260 per pixel in dispersion direction in the blue MODS channel and about 370 in the red channel.

The spectra were reduced using customized ESO-MIDAS scripts. The eight read-out amplifiers per CCD cause an odd/even column pattern resulting from the two read-out amplifiers per quadrant, each having its own bias and gain level. The corresponding bias values were extracted from the overscan regions; a flat field correction was done by flats taken on the same day as the science frames using gas lamps. We averaged 30 flat field exposures to create a master flat. 

\begin{figure}
   \centering
   \includegraphics[height=\hsize,angle=270]{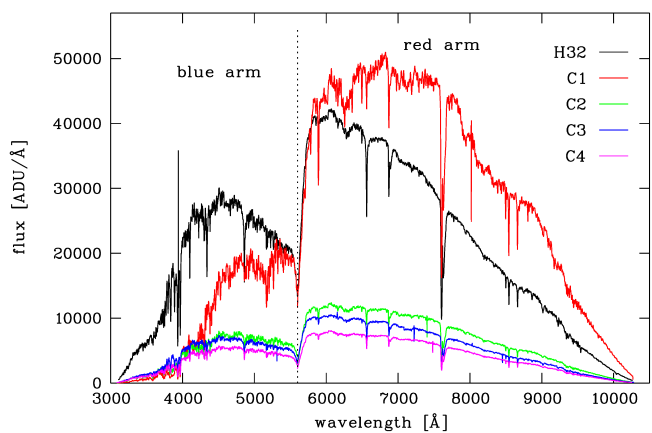}
      \caption{Example spectra of a single exposure of HAT-P-32 and the four comparison stars. The wavelength separation of the blue and the red arm of MODS at 5600~\AA{} is indicated by a vertical dotted line. The features at $\sim$\,3940~\AA{} in the spectrum of HAT-P-32 (black) and at $\sim$\,8011~\AA{} in the spectrum of C1 (red) are caused by CCD errors.}
         \label{plot_examspec_H32}
   \end{figure}

The dispersion direction of the spectra is misaligned to the pixel rows of the detector by $\sim$\,18~pixel in spatial direction over the $\sim$\,2800~pixel in dispersion direction. We rebinned the spectra for alignment. A similar rebinning in dispersion direction was not necessary because no bending was detected over the rather short extent of the slit in spatial direction. Wavelength calibration spectra were obtained with a different slit mask. The slit positions and lengths were the same as in the science mask, whereas the width was reduced to 1\,$''$ to avoid impractically broad lines. The wavelength calibration included a shift in dispersion to correct for the  pixel drift, which was measured by the centroid movements of the telluric O$_2$ lines. According to the wavelength solution of the arc lamp exposures and the scatter of the residual line positions in wavelength, the accuracy of the wavelength calibration is better than 0.3~\AA. One-dimensional spectra were extracted by a simple flux sum in 
spatial direction within a FWHM-dependent aperture. Examples of extracted spectra of HAT-P-32 and the comparison stars are shown in Figure~\ref{plot_examspec_H32}.

To create light curves of the transit event, we integrated the spectral flux for the target and comparison stars in wavelength channels, and performed differential photometry per channel. We compared light curves by using the optimal extraction technique of \citet{Horne86} for the extraction of one-dimensional spectra, and in agreement with \cite{MallonnH19}, we found a more robust spectrophotometry for the simple flux sum. We tested which object aperture and which width of sky stripes minimizes the rms of the light curve residuals after a transit model subtraction using literature transit parameters. This resulted in an aperture of 5.5 times the spatial FWHM in the blue arm and 6.5 times the spatial FWHM in the red arm. We note that these apertures include the flux of the faint companion of HAT-P-32 at 2.9\,$''$ distance. We discuss its influence in detail in Section~\ref{Chap_H32_Mdwarf}. Interestingly for the sky subtraction, we did not find that the widest sky stripes  gave the lowest scatter in the 
light curves. Instead, rather small stripes of 10 pixel width on both sides placed as far away from the object as possible yielded the best quality despite  their higher photon noise. Our explanation is that wider sky stripes need to approach the spectral centroid more closely because of the limited spatial extension of the slits. Thus, wide sky stripes contain more target light of the wide spectral profile wings resulting in an erroneous sky determination.

\begin{table}
\caption{Target HAT-P-32 and the four comparison stars. The table gives the identifier of the star as used in Figure~\ref{ima_aqui_H32}, its GSC name, and the brightness difference $\Delta$V and $\Delta$R to HAT-P-32 in magnitudes.}
\label{H32_compstars}
\begin{center}
\begin{tabular}{cccc}
\hline
\hline
\noalign{\smallskip}
identifier     &  star &  $\Delta$V (mag) &  $\Delta$R (mag) \\
\hline
H32 &HAT-P-32 & & \\
1 & GSC 03281-00957 & 0.09 & -0.24 \\
2 & GSC 03281-00900 & 1.36 & 1.30 \\
3 & GSC 03281-00464 & 1.49 & 1.48 \\
4 & GSC 03281-00640 & 1.76 & 1.78 \\
\hline
\noalign{\smallskip}
\hline
\end{tabular}
\end{center}

\end{table}

\subsection{Photometric monitoring}

A long-term photometric monitoring campaign of the host star HAT-P-32 was carried out with the robotic twin telescope STELLA \citep{Strassmeier2004} and its wide field imager \mbox{WiFSIP} \citep{Weber2012}. The imaging instrument consists of a 4k$\times$4k back-illuminated CCD with a plate scale of 0.322\,$''$/pixel and four read-out amplifiers. It covers a field of 22\,$'\times$ 22\,$'$ on the sky. We monitored HAT-P-32 between December 2011 and March 2013 in the filters Johnson B and Sloan r' and obtained a total of 2456 and 2442 images, respectively. The images were taken in blocks of three frames in B plus three frames in r' with an exposure time of 15~s each. If the conditions allowed, multiple blocks were taken each night. In total, the telescope observed HAT-P-32 on 180 individual nights. We read out the full FoV and applied no defocus.

The data reduction was done with the customized routines  used previously for the long-term photometry of HAT-P-12 and HAT-P-19 in \cite{MallonnH12} and \cite{MallonnH19}. The scripts are written in ESO-MIDAS and call the aperture photometry software SExtractor \citep{Bertin96}. A selection of seven comparison stars in r' and 11 stars in B was found to minimize the scatter in the differential target light curve. We verified that the Lomb-Scargle periodograms (Section~\ref{chap_res_lotephot}) are independent of the choice of the comparison stars. Data points with a target peak count rate higher than 60\,000~ADU, a flux rate lower than 30\% of average, a sky background higher than 1000 ADU, a FWHM higher than 15 pixel, and a semi-major and semi-minor axis ratio of the PSF shape of more than 2.5 were excluded. Finally we averaged the three exposures taken per block, which left us with 480 data points in Johnson B with a point-to-point scatter of 3.4~mmag, and 512 data points in Sloan r' with a scatter of 2.
7~mmag.

\section{Results of the monitoring program and starspot correction}
\label{chap_res_lotephot}

Brightness inhomogeneities on the stellar surface, for example due to starspots \citep{Strassmeier2009}, modify the transit parameters. If the spots are not occulted by the disk of the planet, they lower the effective size of the stellar disk. Therefore, the transit appears deeper than with a homogeneous photosphere \citep{Sing2011, Pont2013}. If spots are located along the transit chord, the occultation causes a flux rise in the measurement and results in a bump in the photometric transit light curve. A fit of a symmetric transit model would yield an underestimated transit depth. \cite{Hartman2011} found a substantial RV jitter of about 80~m$\,$s$^{-1}$ of the host star, and provide arguments why this jitter is more likely to be caused by temporally changing convective inhomogeneities than by starspots. For example, there is a lack of photometric out-of-transit variations in the HATNet data. We monitored the host star in B and r' before and after the transit observation to verify whether photometric 
variations are indeed absent. 

\begin{figure}
   \centering
   \includegraphics[height=\hsize,angle=270]{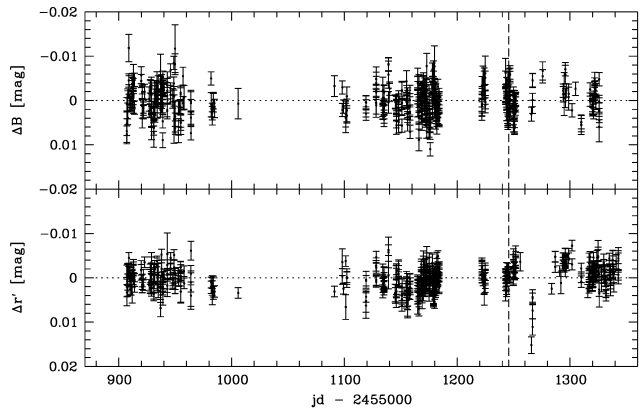}
      \caption{Monitoring light curve of the host star HAT-P-32 observed with STELLA/WiFSIP in Johnson B and Sloan r'. A vertical dashed line indicates the moment of the MODS transit observation.}
         \label{H32_lotephot}
   \end{figure}

\begin{figure}
   \centering
   \includegraphics[height=\hsize,angle=270]{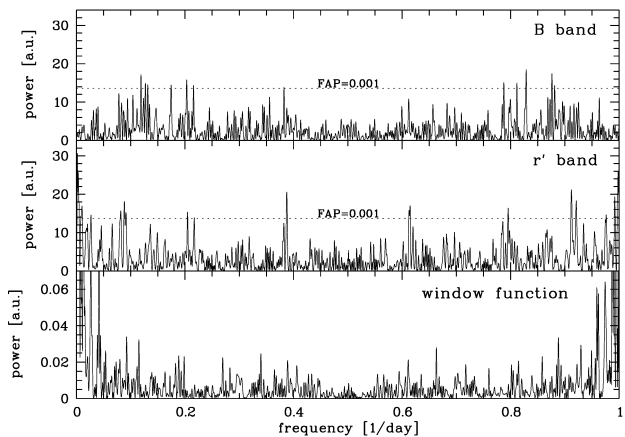}
      \caption{Lomb-Scargle periodogram of the monitoring light curves in B and r'. The dotted horizontal lines indicate the value of FAP\,=\,0.001. At the bottom, an average of the two corresponding window functions is given.}
         \label{H32_periodo}
   \end{figure}

The light curves in both filters after the subtraction of a transit model are shown in Figure~\ref{H32_lotephot}. No significant variability is visible by eye. The photometric brightness at the moment of the transit observations matches the overall mean of the time series. We computed a Lomb-Scargle periodogram (Figure~\ref{H32_periodo}) to search for periodic variations. There are several peaks above the value of a false alarm probability (FAP) of 0.001, e.g., both data sets show a common frequency of $\sim$\,0.2 and $\sim$\,0.39~cycles\,day$^{-1}$. Sine fits to these frequencies reveal semi-amplitudes of 1~mmag and less in all cases. A plain least-squares periodogram used to estimate the amplitude of periodic sine and cosine functions over a wide frequency range confirms that no semi-amplitudes larger than about 1 mmag are present in the data. According to Eq.~4 in \cite{Sing2011}, the correction on the transit depth of HAT-P-32b for a flux variation at the 1\,mmag level is below $1 \times 10^{-4}$. This 
is smaller than our mean uncertainty on $k$ by about a factor of three. Therefore, we conclude that a correction of the derived transmission spectrum for brightness inhomogeneities on the stellar surface is not necessary.

\section{Third-light contribution for the M dwarf companion}
\label{Chap_H32_Mdwarf}

\cite{Adams2013} found a companion object only 2.9\,$''$ to HAT-P-32, which is fainter in the Johnson K band by 3.4 magnitudes. This magnitude difference corresponds to a light contribution of the nearby object (third light) of $\sim$\,4.5\% in K. Third light has the opposite effect to unocculted starspots on the transit parameters and dilutes the transit depth. A third-light contribution of 4.5\% in the K band would damp the true transit depth by the same amount of 4.5\%.

\begin{figure}
   \centering
   \includegraphics[height=\hsize,angle=270]{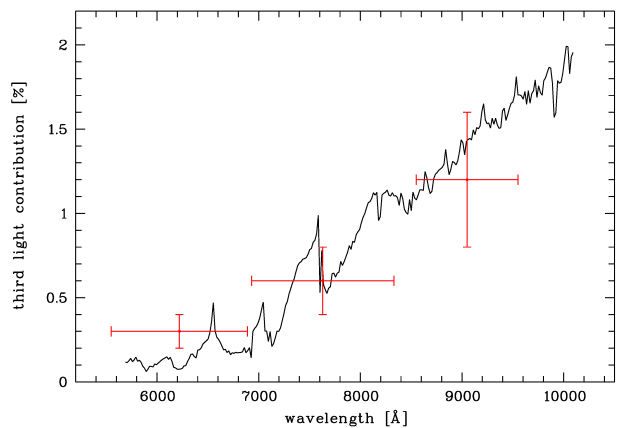}
      \caption{Third-light contribution of the nearby M dwarf to the integrated light of HAT-P-32. The broad-band measurements of \cite{Zhao2014} in r', i', and z' are plotted in red. }
         \label{plot_thirdlight}
   \end{figure}

The light of this companion is included in our aperture. We approximate the third-light contribution $\Delta f$ with the ratio of the peak intensities of the spectral profiles of both objects. Because the peak of the M dwarf is influenced by the wings of the PSF of HAT-P-32, we removed HAT-P-32 from the spectral profile. The one side of the profile opposite the M dwarf was mirrored and subtracted from the other side of the profile containing the companion profile. The peak intensity ratio was calculated individually for each exposure of the time series. Because no significant time dependence could be found (potentially caused by variability of the M dwarf), the values were averaged. We estimated the third-light contribution only from the red MODS data, because at blue wavelengths ($\lambda\,<\,5600~\AA{}$) the influence of the M dwarf is negligible. A plot of this intensity ratio over wavelength in steps of $\sim$\,15~\AA{} is presented in Figure~\ref{plot_thirdlight}. Many spectral absorption features typical of an M dwarf are visible, for example  Na, H$\alpha$, FeH, TiO, and VO. Our derived values are in very good agreement with previous estimations \citep[G13;][]{Hartman2011,Adams2013,Zhao2014}, but provide a higher spectral resolution.  At the reddest light curves of our sample, the transit depth needs to be corrected by almost 2\%, which is about the size of the transit depth uncertainties. The uncertainty of the third-light contamination of about 0.05\% is small enough to be negligible in the error budget of the transit parameters. Stellar properties of the M dwarf companion were derived by \cite{Zhao2014} and \cite{Ngo2015} and it is out of the scope of the current work to refine them.

\section{Transit light curve analysis and results}

\subsection{Creation of light curves}
We created a set of light curves over wavelength by the integration of the spectral flux in multiple wavelength bands. Each band yields one data point in the planetary transmission spectrum and the width of these bands determines the achievable spectral resolution. However, the bands need to be wide enough to collect sufficient flux for a S/N$\gtrsim$1000 per data point in the spectrophotometric light curves, an accuracy necessary to obtain high-precision transit parameters. We use 62 wavelength bands with an average width of 100~\AA{} with narrower widths at the wavelengths of Na, K, H$\alpha$, which are potential absorption lines in the planetary atmosphere \citep{Charbonneau2002,Sing2011b,Jensen2012}. The spectral resolution of the planetary spectrum is $\mathrm{R} \sim 60$. The specific borders of the bands are chosen to exclude faulty CCD pixels and fully include wide absorption features in the stellar spectra like the Balmer lines. We note that partially covered absorption features reduce the 
spectrophotometric accuracy because the features move in wavelength at a sub-\AA{} level from exposure to exposure following the object centroid jitter in the wide slit. 

\begin{figure}
   \centering
   \includegraphics[height=\hsize,angle=270]{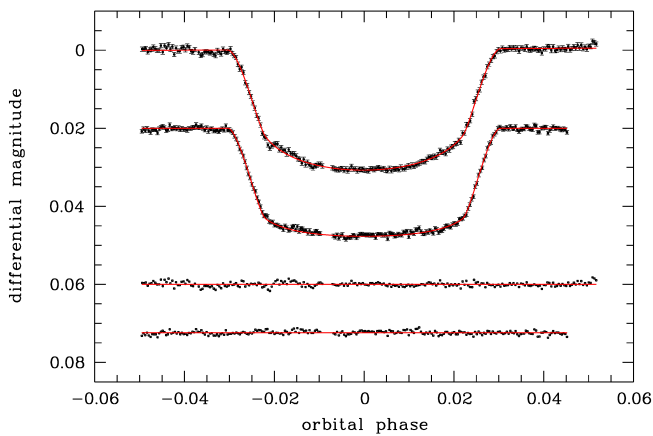}
      \caption{Averaged light curves for the blue and the red arm, and their corresponding residuals (from top to bottom). Overplotted in red are the best-fit transit models including a linear (blue arm) and a quadratic (red arm) detrending function over time.}
         \label{plot_blue_red}
   \end{figure}

We chose the combination of the four comparison stars in flux that provided the light curve with the lowest photometric scatter after a subtraction of a transit model using literature transit parameter. The obtained optimal combination was found to be wavelength dependent. Because of this wavelength dependence we did not produce a white light curve by an integration of the spectral flux over the entire spectral range. Instead, we first produced the set of 62 light curves over wavelength, and then formed their weighted average with the point-to-point scatter as inverse weighting factor. Because the exposures of the blue and the red CCD have slightly different sampling rates, we formed a combined ``blue'' light curve of the 19 wavelength channels of the blue MODS arm and a ``red'' light curve from the 43 light curves of the red arm. Both averaged light curves are shown in Figure~\ref{plot_blue_red}, while the set of 62 light curves over wavelength is shown in Figure~\ref{plot_all_lcs}. The blue and red light 
curves have a point-to-point scatter (rms) of 0.56 and 0.44~mmag, which are respectively a factor of about 3.9 and 3.6  higher than the photon noise of the target and the comparison star combination. The 62 light curves show a scatter from 0.83 to 3.7~mmag (mean 1.2~mmag), which is a factor of 1.3 to 2.8 higher than the photon noise (mean 1.9). These results are similar to the values reached with GEMINI-N/GMOS in comparable spectrophotometric studies \citep[G13;][]{Stevenson2014}. The properties of the chromatic light curves are summarized in Table \ref{tab_62lcs}.

  \onlfig{
   \begin{figure*}
   \centering
   \includegraphics[width=16.5cm]{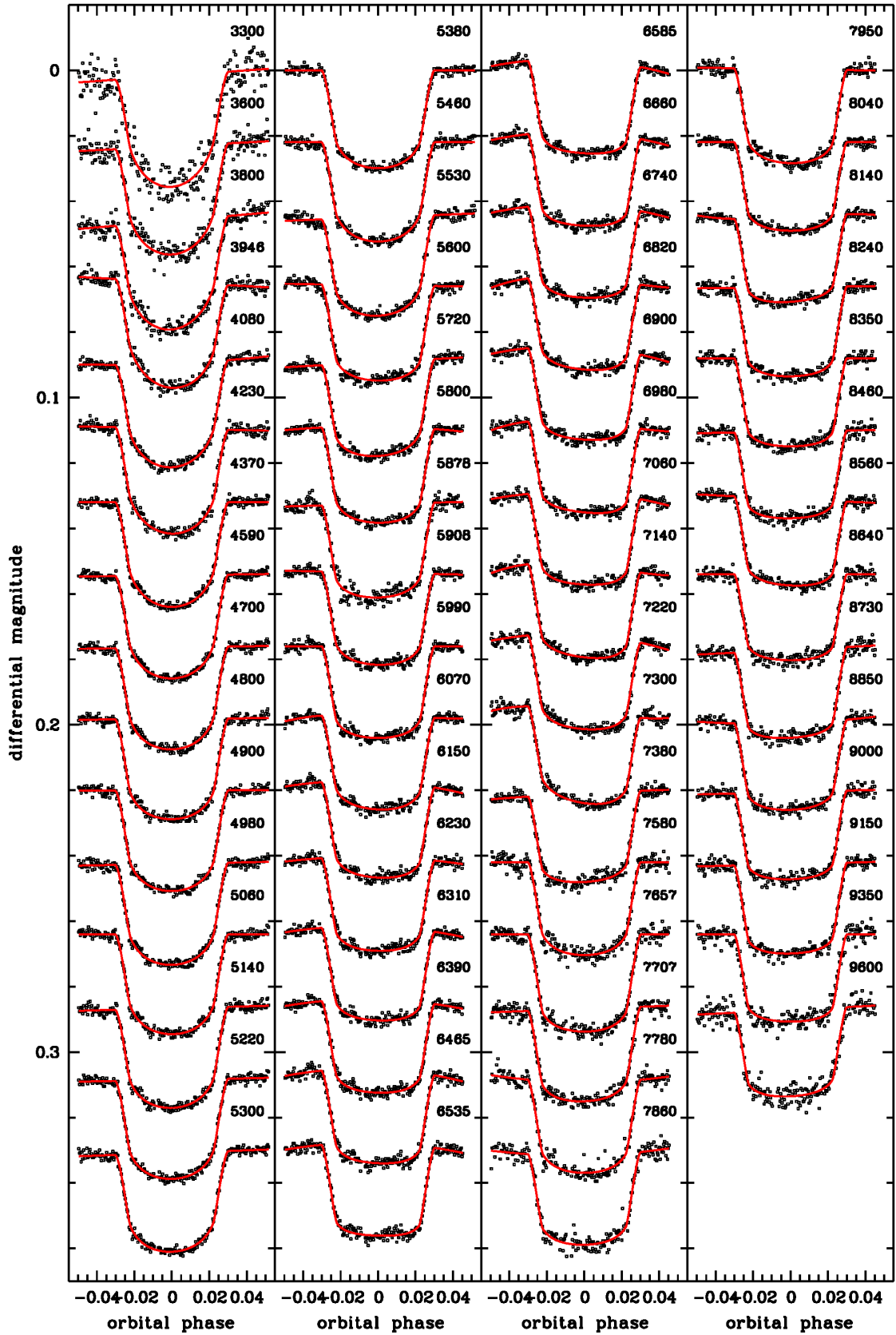}
      \caption{Set of 62 light curves over wavelength offset for clarity. Overplotted in red are the models of the transit light curve including the detrending polynomial.}
         \label{plot_all_lcs}
   \end{figure*}
  }

  \onlfig{
   \begin{figure*}
   \centering
   \includegraphics[width=16.5cm]{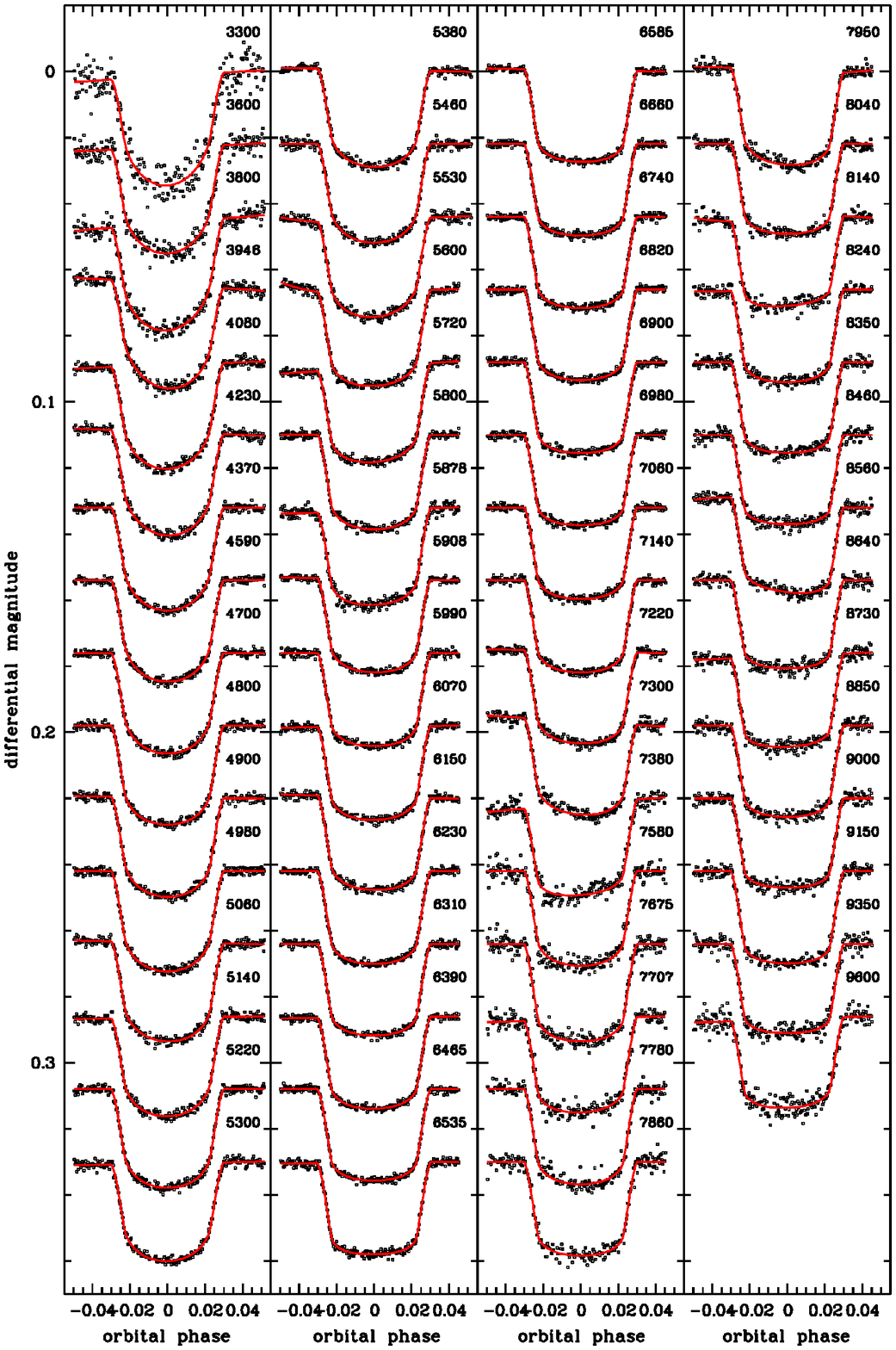}
      \caption{Similar to Figure \ref{plot_all_lcs}, but after the chromatic mode correction.}
         \label{plot_all_lcs_cmc}
   \end{figure*}
  }

\subsection{Light curve modeling}
\label{chap_lc_mod}

All transit light curves in this work were analyzed with the publicly available software \mbox{JKTEBOP}\footnote{http://www.astro.keele.ac.uk/jkt/codes/jktebop.html} \citep{Southworth2004,Southworth2008} in version 34. The transit fit parameters consist of the sum of the fractional planetary and stellar radius, $r_{\star} + r_p$, their ratio $k=r_p/r_{\star}$, the orbital inclination $i$, the transit midtime $T_0$, and the host star limb-darkening coefficients (LDC) $u$ and $v$ of the quadratic limb-darkening law. The index ``$\star$'' refers to the host star and ``p'' refers to the planet. The dimensionless fractional radius is the absolute radius in units of the orbital semi-major axis $a$; $r_{\star} = R_{\star}/a$ and $r_p = R_p/a$. In addition, \mbox{JKTEBOP} allows  the simultaneous fit of polynomials and sine curves for detrending of the light curves. We used the Bayesian Information Criterion \citep[BIC,][]{Schwarz78} to choose the detrending models for each light curve of the chromatic set and for 
the blue and red light curves. The BIC consistently favored low-order polynomials over time from order zero to three. The blue light curve showed a very weak correlation of the residuals and the FWHM of the spectral spatial profile. However, the BIC value did not justify the inclusion of a FWHM-dependent term in the detrending function. No other correlations between light curve residuals and external parameters like pixel position or rotation angle was found. Table \ref{tab_62lcs} lists the order of the time-dependent polynomials used.

Throughout this work, we employed the quadratic limb-darkening law. We followed the common approach of fixing the quadratic coefficient $v$ to its theoretical value, while fitting for the linear coefficient $u$ \citep[e.g.,][]{Mancini2014,Southworth2015,MallonnH12}. In the error analysis, $v$ is perturbed to account for systematic errors in its theoretical calculation. Theoretical values for $v$ were obtained from \cite{Claret2013} with sufficient spectral resolution (Claret, priv. comm.). The stellar parameters used were the ones from \cite{Hartman2011} for the case of a circular planetary orbit.

We employed a constant photometric uncertainties over time per light curve because of the stable flux level (see Figure~\ref{plot_obs_cond}) and adopted the standard deviation of the residuals after an initial transit fit. We further enlarged the photometric uncertainties by the so-called $\beta$ factor, a concept introduced by \cite{Gillon06} and \cite{Winn08}, to account for the presence of correlated noise in the light curves. The $\beta$ factor used here is the mean of the ratio of the standard deviation of the residuals and the standard deviation of white noise when binned in intervals from 10 to 30 minutes.

The estimation of the transit parameter uncertainties was done with  ``task 8'' in \mbox{JKTEBOP} \citep{Southworth2005}, which is a Monte Carlo simulation, and with ``task 9'' \citep{Southworth2008}, which is a residual-permutation algorithm that takes into account correlated noise. We ran the Monte Carlo simulation with 5000 steps. As final parameter uncertainties we adopted the larger value of both methods.

\subsection{Results for the blue and red light curve}

First, we analyzed the blue and the red light curve. After an initial fit using literature transit parameters, we estimated the photometric uncertainties, enlarged them by the $\beta$ factor (see  Section~\ref{chap_lc_mod}) and removed outliers in the light curve residuals by a 3-$\sigma$ clipping. We calculated the BIC value to estimate the order of the detrending polynomial over time, which resulted in a first-order polynomial for the blue light curve (two free parameters $c_0$ and $c_1$) and a second-order polynomial for the red light curve (three free parameters $c_0$, $c_1$, and $c_2$). Then, we performed our transit light curve fit with the free parameters $r_{\star} + r_p$, $k$, $i$, $T_0$, $u$, and $c_{0,1}$ respectively $c_{0,1,2}$. The orbital period was fixed to the value of \cite{Seeliger2014}, the orbital eccentricity was fixed to zero following \cite{Zhao2014}. The results of the transit parameters for the blue and red light curve are summarized in Table~\ref{transitparam_H32}. The value of $k$ 
of the red light curve is corrected by the flux weighted average of the third-light contribution from the nearby M dwarf (see Section \ref{Chap_H32_Mdwarf}) using Eq.~4 in \cite{Sing2011}. A comparison to previous results from \cite{Hartman2011}, G13, and \cite{Seeliger2014} reveals agreement to within 1\,$\sigma$. The MODS transit corresponds to epoch 849 with regard to the ephemeris zero point defined in \cite{Hartman2011}. Our derived values for T$_0$ agree to within 1\,$\sigma$ of the calculated value using the period of \cite{Seeliger2014}.

\begin{table*}
\small
\caption{ Derived transit parameters of this work in comparison to literature values for HAT-P-32b.}
\label{transitparam_H32}
\begin{center}
\begin{tabular}{lccccc}
\hline
\hline
\noalign{\smallskip}
                   &  this work, blue light &  this work, red light &  \cite{Hartman2011} & G13 & \cite{Seeliger2014} \\
\hline
\noalign{\smallskip}
$a/R_{\star}\,=\,1/r_{\star}$ &    6.047 $\pm$ 0.087       &   6.046 $\pm$0.055       & 6.05$^{+0.03}_{-0.04}$  &  6.091$^{+0.047}_{-0.036}$  & 6.060 $\pm$ 0.009   \\
$k\,=\,r_p/r_{\star}$   &    0.1515 $\pm$ 0.0012     &   0.1505\tablefootmark{a} $\pm$ 0.0005   & 0.1508 $\pm$ 0.0004     &  0.1515 $\pm$ 0.0012       &  0.1507 $\pm$ 0.0004 \\
i (deg)           &    88.61 $\pm$ 0.84        &   88.56 $\pm$ 0.57       & 88.9 $\pm$ 0.4          &  89.12$^{+0.68}_{-0.61}$   &  88.98 $\pm$ 0.10   \\
T$_0$ - 2456245 (BJD$_{\mathrm{TDB}})$   &  0.80334 $\pm$ 0.00014  & 0.80349  $\pm$ 0.00008& & & \\
$u$               & 0.464 $\pm$ 0.031 & 0.216 $\pm$ 0.018 & & & \\
$v$               & 0.212 & 0.224 & & & \\
rms (mmag)        & 0.56 & 0.44 & & & \\
N$_{data}$        & 238 & 240 & & & \\
$\beta $          & 1.98 & 1.47 & & & \\
\hline
\end{tabular}
\end{center}
\tablefoot{
\tablefoottext{a}{This value is corrected for the third-light contribution of the M dwarf companion, see Section~\ref{Chap_H32_Mdwarf}.}
}
\end{table*}

\subsection{Results for the set of 62 chromatic light curves}

The main goal of the current work is the derivation of a planetary transmission spectrum in low spectral resolution from the near-UV to the near-IR. The transit parameters $r_{\star}$, $i$, and T$_0$, which are expected to be independent of wavelength, were fixed in the transit fit. For comparability with the transmission spectrum presented by G13, we used their values for $r_{\star}$ and $i$; T$_0$ was set to the prediction of the ephemeris of \cite{Seeliger2014}. Free parameters per light curve are $k$, $u$, and the coefficients of the detrending polynomial over time. The uncertainty of the wavelength-independent parameters are assumed to affect all chromatic light curves equally, hence they can be neglected in a differential search for variations in the planetary radius \citep{Mancini2014,Southworth2015}. Our uncertainties on $k$ are therefore relative uncertainties.

\begin{figure*}
   \centering
   \includegraphics[height=\hsize,angle=270]{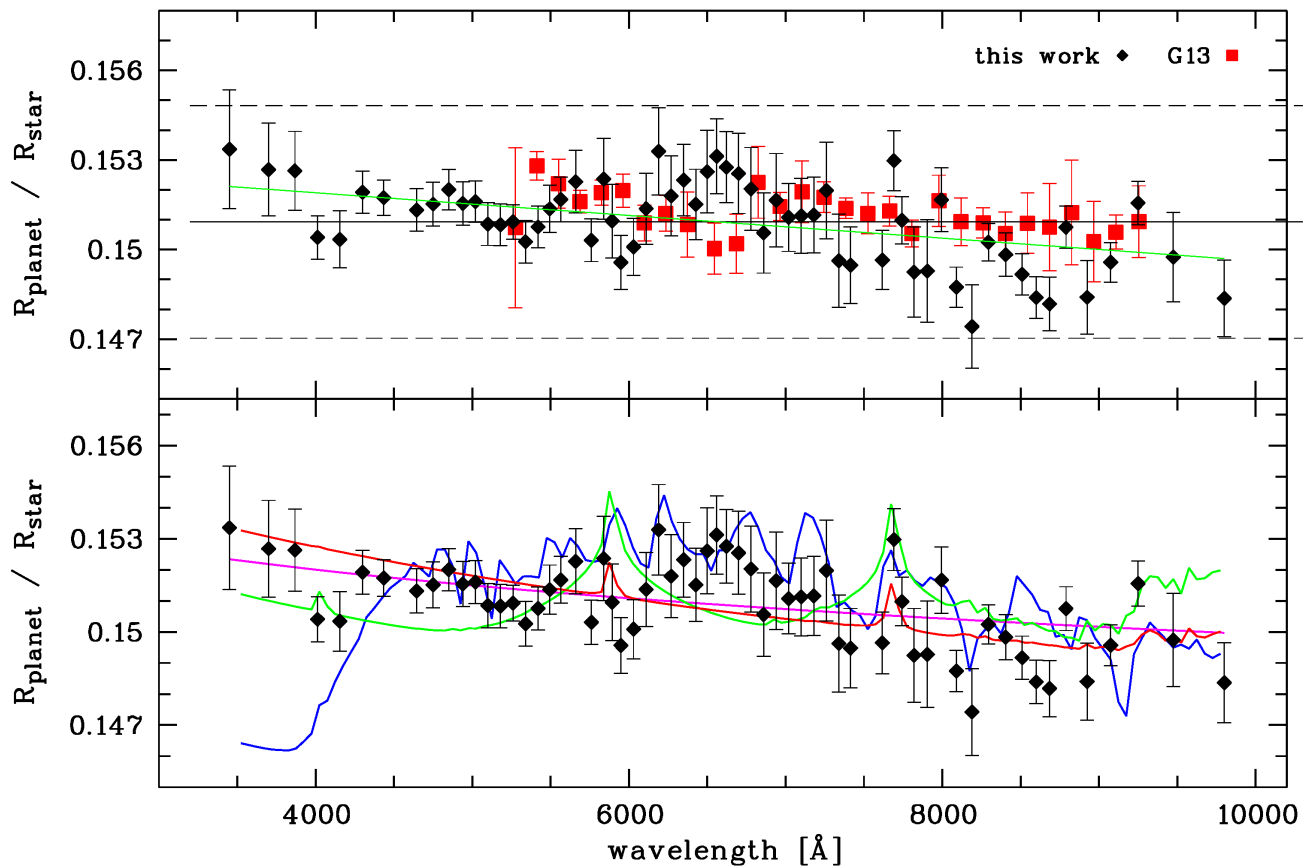}
      \caption{Transmission spectrum of HAT-P-32b derived from the MODS transit measurement. Upper panel: The transmission spectrum of this work (black) is compared to the derived spectrum of G13 (red). Overplotted in green is the best fit of a linear function over wavelength. The horizontal dashed lines indicate three pressure scale heights of the planetary atmosphere. Lower panel: Our spectrum is overplotted with a cloud-free solar-composition model of 1750~K (blue line), a cloud-free solar composition model of 1750~K without TiO (green line), a solar composition model of 1750~K including H2 Rayleigh scattering increased by a factor of 100 (red line), and a best-fit Rayleigh scattering slope of 890~K (magenta line). }
         \label{plot_transspec_2pan}
   \end{figure*}

All transit values of $k$ of the light curves redder than 5600~\AA{} were corrected for the third-light contribution from the nearby M dwarf (see Section \ref{Chap_H32_Mdwarf}) using Eq.~4 in \cite{Sing2011}. The resulting transmission spectrum is given in Table~\ref{tab_62lcs} and Figure~\ref{plot_transspec_2pan}. It is broadly consistent with the planetary spectrum derived by G13, with individual data points deviating by about 2.5\,$\sigma$ at 5430~\AA{}, 6560~\AA{}, and 8090~\AA{}. We compared our measurements with several theoretical \cite{Fortney2010} models computed for 1750~K atmospheric temperature and scaled to the planetary parameters of HAT-P-32b. A cloud-free solar-composition model dominated by TiO absorption gives a low-quality fit of $\chi^2$ of 212.1 for 61 degrees of freedom (DOF); a solar-composition model with TiO artificially removed results in $\chi^2\,=\,168.9$ (DOF=61); and a solar composition model without TiO and with a Rayleigh scattering component with a cross section 100$\times$ 
that of H$_2$ gives $\chi^2\,=\,90.7$ (DOF=61). For all three models the corresponding one-tailed probabilities $P$ of a $\chi^2$ test are below 0.01, i.e., these models are significantly ruled out by the data. Also ruled out is a wavelength-independent absorption as a flat line fitted to the data results in $\chi^2\,=\,111.2$ (DOF=61). An acceptable fit can be achieved by a linear function over wavelength, resulting in $\chi^2\,=\,79.9$ (DOF=60).

\subsubsection{Common mode correction}

The residuals of the blue and the red light curve in Figure~\ref{plot_blue_red} show correlated noise at the sub-mmag level. We made the assumption that this correlated noise pattern is common for all chromatic light curves and tested whether its subtraction from all chromatic light curves can improve the derived transmission spectrum. Because the residuals were subtracted equally from all chromatic light curves, this correction does not affect the differential variations of the planetary radius. However, we found this common mode correction to have little effect on the light curves and the resulting transmission spectrum: on average it does not reduce the rms of the chromatic light curves and decreases the $\beta$ factor by about 15\%. The attempted common mode correction does not affect the detrending. 

\begin{figure}
   \centering
   \includegraphics[height=\hsize,angle=270]{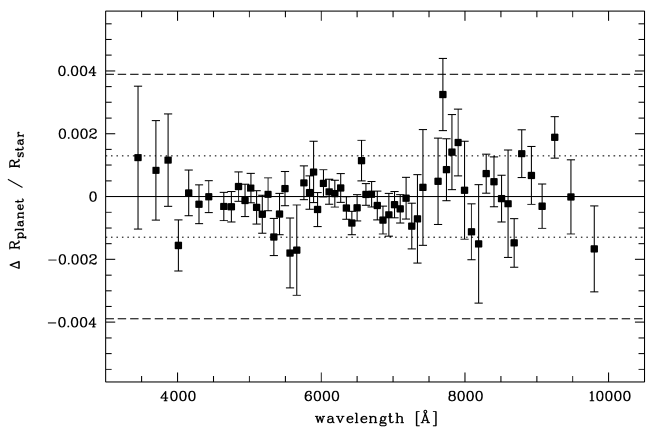}
      \caption{Transmission spectrum of HAT-P-32b after the chromatic mode correction. The horizontal dotted and dashed lines indicate respectively one and three pressure scale heights of the planetary atmosphere.}
         \label{plot_transspec_cmc}
   \end{figure}

However, we see that the detrending has a great influence on the uncertainties of $k$. Light curves using a more flexible detrending model of order two and three have on average more than 50 percent higher uncertainties on $k$ than models using a polynomial of order zero or one. A common subtraction of the trend of the red light curve from all chromatic light curves of the red camera would also not yield an improvement  because a wavelength dependence of the trends can be seen by eye in Figure~\ref{plot_all_lcs}. Instead, we tested whether a correction of trends common to neighboring wavelength channels (here called chromatic mode correction) can clean the individual light curves on average. First, from each chromatic light curve we subtracted a transit model fixed to the mean of the parameters of the blue and red light curve with tabulated LDC and without a detrending function. As a result we obtained the chromatic light curve residuals per wavelength channel still including the individual trend. For each 
chromatic transit light curve, we averaged and subtracted the residuals of the eight neighboring channels (four toward the blue and four toward the red side). A new calculation of the BIC values showed that this new version of the chromatic light curves needed substantially lower orders of the polynomial over time with orders higher than one being an exception. Furthermore, the rms of the point-to-point variation decreased on average by $\sim$\,10\%, and the $\beta$ factor by $\sim$\,30\%. The uncertainties in $k$ decreased on average by $\sim$\,30\%. The corresponding chromatic mode corrected light curves are shown in Fig. \ref{plot_all_lcs_cmc}, their transmission spectrum in Figure \ref{plot_transspec_cmc}, and a summary of the light curve properties is given in Table \ref{tab_62lcs_cmc}. We note that this chromatic mode corrected transmission spectrum no longer holds information about overall gradients  because they would have been subtracted by residuals versus a transit model of constant $k$. Instead, 
this transmission spectrum exhibits potential $k$ variations regarding the adjacent channels and is therefore  useful in the  search for absorption in the cores of, e.g., Na and K.
 
The standard deviation of the data points over our entire wavelength range is 0.8 atmospheric pressure scale heights of HAT-P-32b. For the 39 measurements between 4000~\AA{} and 7400~\AA{}, we achieve a standard deviation of 0.52 scale heights. One scale height amounts to $\sim$\,1000~km using the stellar and planetary parameters of \cite{Hartman2011} for zero eccentricity. We find an increased absorption of 2.8\,$\sigma$ in the wavelength channel of potassium compared to the neighboring channels. No increased absorption was detected in the wavelength channels of Na and H$\alpha$.

\onltab{
\begin{table*}
\small
\caption{Characteristics of the 62 chromatic light curves and the derived values of $k$ with relative uncertainties. }
\label{tab_62lcs}
\begin{center}
\begin{tabular}{lrccccccc}

\hline
   wavelength  &     width  &  comparison  &  rms   & $\beta$ & poly  & $u$               & $v$  & $k$ \\
  (\AA{})      &   (\AA{})  &    star      & (mmag) &        &        &                   &      &     \\
\hline
\hline
\noalign{\smallskip}
               &            &              &        &        &        &                    &            &     \\  
 3300 - 3600   &       300  &  2,4         &  3.68  &  1.10  &   1    &  0.676 $\pm$ 0.052 &  0.096   & 0.15335 $\pm$ 0.00198  \\
 3600 - 3800   &       200  &  1,2,3,4     &  2.24  &  1.35  &   1    &  0.612 $\pm$ 0.040 &  0.141   & 0.15268 $\pm$ 0.00155  \\
 3800 - 3937   &       137  &  2,3,4       &  1.70  &  1.43  &   1    &  0.653 $\pm$ 0.032 &  0.080   & 0.15263 $\pm$ 0.00132  \\
 3946 - 4080   &       134  &  2,3,4       &  1.31  &  1.06  &   1    &  0.638 $\pm$ 0.028 &  0.124   & 0.15040 $\pm$ 0.00072  \\
 4080 - 4230   &       150  &  1,2,3,4     &  0.89  &  1.21  &   2    &  0.562 $\pm$ 0.025 &  0.182   & 0.15034 $\pm$ 0.00095  \\
 4230 - 4370   &       140  &  1,3         &  1.20  &  1.22  &   1    &  0.580 $\pm$ 0.020 &  0.118   & 0.15192 $\pm$ 0.00070  \\
 4370 - 4500   &       130  &  1,2,3,4     &  0.85  &  1.26  &   0    &  0.545 $\pm$ 0.025 &  0.175   & 0.15173 $\pm$ 0.00059  \\
 4590 - 4700   &       110  &  1,2,3       &  0.83  &  1.55  &   1    &  0.498 $\pm$ 0.028 &  0.223   & 0.15132 $\pm$ 0.00071  \\
 4700 - 4800   &       100  &  1,2         &  0.93  &  1.46  &   1    &  0.460 $\pm$ 0.029 &  0.231   & 0.15151 $\pm$ 0.00073  \\
 4800 - 4900   &       100  &  1,2         &  1.00  &  1.31  &   1    &  0.391 $\pm$ 0.020 &  0.251   & 0.15200 $\pm$ 0.00067  \\
 4900 - 4980   &        80  &  1,2,3       &  0.90  &  1.43  &   1    &  0.432 $\pm$ 0.029 &  0.224   & 0.15154 $\pm$ 0.00072  \\
 4980 - 5060   &        80  &  1,2,3       &  0.96  &  1.30  &   1    &  0.439 $\pm$ 0.028 &  0.228   & 0.15160 $\pm$ 0.00069  \\
 5060 - 5140   &        80  &  1,2,3       &  0.99  &  1.30  &   0    &  0.434 $\pm$ 0.029 &  0.228   & 0.15084 $\pm$ 0.00071  \\
 5140 - 5220   &        80  &  1,2,3,4     &  1.01  &  1.26  &   1    &  0.438 $\pm$ 0.028 &  0.206   & 0.15083 $\pm$ 0.00070  \\
 5220 - 5300   &        80  &  1,2,3       &  0.84  &  1.01  &   1    &  0.414 $\pm$ 0.022 &  0.221   & 0.15092 $\pm$ 0.00057  \\
 5300 - 5380   &        80  &  1,2,3       &  0.90  &  1.39  &   1    &  0.424 $\pm$ 0.028 &  0.230   & 0.15026 $\pm$ 0.00071  \\
 5380 - 5460   &        80  &  1,2,3,4     &  0.86  &  1.05  &   0    &  0.383 $\pm$ 0.024 &  0.237   & 0.15075 $\pm$ 0.00069  \\
 5460 - 5530   &        70  &  1,2,3       &  0.97  &  1.42  &   0    &  0.406 $\pm$ 0.021 &  0.218   & 0.15136 $\pm$ 0.00078  \\
 5530 - 5600   &        70  &  1,2,3,4     &  1.09  &  1.52  &   1    &  0.363 $\pm$ 0.025 &  0.243   & 0.15167 $\pm$ 0.00075  \\
 5600 - 5720   &       120  &  1,3,4       &  1.18  &  1.84  &   1    &  0.269 $\pm$ 0.034 &  0.229   & 0.15227 $\pm$ 0.00104  \\
 5720 - 5800   &        80  &  1,4         &  1.03  &  1.45  &   1    &  0.290 $\pm$ 0.023 &  0.240   & 0.15030 $\pm$ 0.00071  \\
 5800 - 5878   &        78  &  1,4         &  0.84  &  1.64  &   2    &  0.284 $\pm$ 0.021 &  0.239   & 0.15236 $\pm$ 0.00135  \\
 5878 - 5908   &        30  &  1,4         &  1.66  &  1.57  &   1    &  0.254 $\pm$ 0.041 &  0.237   & 0.15095 $\pm$ 0.00123  \\
 5908 - 5990   &        82  &  1,4         &  1.04  &  1.81  &   1    &  0.263 $\pm$ 0.029 &  0.243   & 0.14956 $\pm$ 0.00090  \\
 5990 - 6070   &        80  &  1,4         &  1.08  &  1.81  &   0    &  0.235 $\pm$ 0.031 &  0.244   & 0.15008 $\pm$ 0.00095  \\
 6070 - 6150   &        80  &  1,4         &  1.02  &  1.36  &   3    &  0.270 $\pm$ 0.022 &  0.234   & 0.15136 $\pm$ 0.00119  \\
 6150 - 6230   &        80  &  1,4         &  1.07  &  1.66  &   2    &  0.253 $\pm$ 0.028 &  0.230   & 0.15328 $\pm$ 0.00145  \\
 6230 - 6310   &        80  &  1,4         &  1.07  &  1.55  &   2    &  0.243 $\pm$ 0.026 &  0.227   & 0.15179 $\pm$ 0.00133  \\
 6310 - 6390   &        80  &  1,4         &  0.99  &  1.54  &   2    &  0.226 $\pm$ 0.024 &  0.236   & 0.15232 $\pm$ 0.00120  \\
 6390 - 6465   &        75  &  1,4         &  1.05  &  1.45  &   2    &  0.240 $\pm$ 0.024 &  0.233   & 0.15151 $\pm$ 0.00117  \\
 6465 - 6535   &        70  &  1,4         &  1.15  &  1.60  &   2    &  0.223 $\pm$ 0.029 &  0.236   & 0.15260 $\pm$ 0.00138  \\
 6535 - 6585   &        50  &  1,3,4       &  1.11  &  1.48  &   2    &  0.148 $\pm$ 0.026 &  0.264   & 0.15312 $\pm$ 0.00125  \\
 6585 - 6660   &        75  &  1,4         &  0.98  &  1.66  &   2    &  0.196 $\pm$ 0.026 &  0.240   & 0.15275 $\pm$ 0.00119  \\
 6660 - 6740   &        80  &  1,4         &  1.15  &  1.54  &   2    &  0.202 $\pm$ 0.028 &  0.232   & 0.15254 $\pm$ 0.00133  \\
 6740 - 6820   &        80  &  1,4         &  1.13  &  1.60  &   2    &  0.212 $\pm$ 0.029 &  0.228   & 0.15202 $\pm$ 0.00138  \\
 6820 - 6900   &        80  &  1,4         &  1.09  &  1.60  &   3    &  0.226 $\pm$ 0.028 &  0.229   & 0.15055 $\pm$ 0.00134  \\
 6900 - 6980   &        80  &  1,3         &  1.18  &  1.87  &   2    &  0.221 $\pm$ 0.036 &  0.232   & 0.15164 $\pm$ 0.00156  \\
 6980 - 7060   &        80  &  1,3         &  1.11  &  1.29  &   3    &  0.184 $\pm$ 0.024 &  0.224   & 0.15107 $\pm$ 0.00114  \\
 7060 - 7140   &        80  &  1,2,4       &  1.06  &  1.59  &   2    &  0.217 $\pm$ 0.027 &  0.227   & 0.15112 $\pm$ 0.00125  \\
 7140 - 7220   &        80  &  1,2,4       &  1.03  &  1.65  &   3    &  0.229 $\pm$ 0.027 &  0.221   & 0.15116 $\pm$ 0.00126  \\
 7220 - 7300   &        80  &  1,2,3,4     &  1.14  &  2.03  &   2    &  0.264 $\pm$ 0.036 &  0.228   & 0.15197 $\pm$ 0.00161  \\
 7300 - 7380   &        80  &  1,2,3,4     &  1.24  &  1.68  &   3    &  0.290 $\pm$ 0.033 &  0.219   & 0.14962 $\pm$ 0.00155  \\
 7380 - 7450   &        70  &  2,3,4       &  1.24  &  1.53  &   2    &  0.198 $\pm$ 0.032 &  0.223   & 0.14947 $\pm$ 0.00127  \\
 7580 - 7675   &        95  &  2,3,4       &  1.60  &  1.26  &   0    &  0.309 $\pm$ 0.032 &  0.221   & 0.14964 $\pm$ 0.00099  \\
 7675 - 7710   &        35  &  2,3,4       &  1.49  &  1.40  &   0    &  0.290 $\pm$ 0.031 &  0.220   & 0.15298 $\pm$ 0.00100  \\
 7710 - 7780   &        70  &  2           &  1.39  &  1.08  &   1    &  0.247 $\pm$ 0.024 &  0.225   & 0.15098 $\pm$ 0.00078  \\
 7780 - 7860   &        80  &  2,3,4       &  1.68  &  1.41  &   2    &  0.257 $\pm$ 0.038 &  0.222   & 0.14924 $\pm$ 0.00151  \\
 7860 - 7950   &        90  &  2           &  1.79  &  1.00  &   2    &  0.227 $\pm$ 0.032 &  0.220   & 0.14927 $\pm$ 0.00171  \\
 7950 - 8040   &        90  &  2,3         &  1.33  &  1.64  &   1    &  0.276 $\pm$ 0.035 &  0.224   & 0.15166 $\pm$ 0.00107  \\
 8040 - 8140   &       100  &  1,2,4       &  0.85  &  1.30  &   0    &  0.223 $\pm$ 0.029 &  0.222   & 0.14874 $\pm$ 0.00067  \\
 8140 - 8240   &       100  &  1,2,3,4     &  0.90  &  1.46  &   3    &  0.161 $\pm$ 0.024 &  0.217   & 0.14742 $\pm$ 0.00139  \\
 8240 - 8350   &       110  &  2,3,4       &  1.07  &  1.01  &   1    &  0.193 $\pm$ 0.028 &  0.220   & 0.15024 $\pm$ 0.00063  \\
 8350 - 8460   &       110  &  2,3,4       &  1.14  &  1.13  &   0    &  0.172 $\pm$ 0.022 &  0.217   & 0.14982 $\pm$ 0.00072  \\
 8460 - 8560   &       100  &  2,3,4       &  1.22  &  1.00  &   1    &  0.160 $\pm$ 0.021 &  0.210   & 0.14916 $\pm$ 0.00069  \\
 8560 - 8640   &        80  &  1,2,3,4     &  0.99  &  1.19  &   1    &  0.181 $\pm$ 0.020 &  0.217   & 0.14839 $\pm$ 0.00070  \\
 8640 - 8730   &        90  &  2,3,4       &  1.29  &  1.24  &   0    &  0.172 $\pm$ 0.028 &  0.215   & 0.14817 $\pm$ 0.00090  \\
 8730 - 8850   &       120  &  2,3,4       &  1.25  &  1.00  &   1    &  0.139 $\pm$ 0.021 &  0.219   & 0.15074 $\pm$ 0.00070  \\
 8850 - 9000   &       150  &  2,3,4       &  1.11  &  1.09  &   2    &  0.184 $\pm$ 0.021 &  0.219   & 0.14839 $\pm$ 0.00123  \\
 9000 - 9150   &       150  &  2,3,4       &  1.13  &  1.01  &   1    &  0.168 $\pm$ 0.029 &  0.214   & 0.14957 $\pm$ 0.00066  \\
 9150 - 9350   &       200  &  2,3,4       &  1.14  &  1.09  &   1    &  0.166 $\pm$ 0.021 &  0.217   & 0.15155 $\pm$ 0.00072  \\
 9350 - 9600   &       250  &  2,3,4       &  1.79  &  1.54  &   0    &  0.171 $\pm$ 0.049 &  0.216   & 0.14974 $\pm$ 0.00149  \\
 9600 - 10000  &       400  &  2,3,4       &  2.20  &  1.00  &   1    &  0.199 $\pm$ 0.037 &  0.209   & 0.14836 $\pm$ 0.00128  \\
\hline                                                                         
\end{tabular}                                                                  
\end{center}
\end{table*}
}

\onltab{
\begin{table*}
\small
\caption{Characteristics of the 62 chromatic light curves after the chromatic mode correction. $\delta\,k$ is the difference in $k$ of each channel compared to the mean of the eight adjacent channels. }
\label{tab_62lcs_cmc}
\begin{center}
\begin{tabular}{lrcccccc}

\hline
   wavelength  &     width   &  rms   & $\beta$ & poly  & $u$               & $v$  & $\Delta\,k$ \\
  (\AA{})      &   (\AA{})   & (mmag) &        &        &                   &      &     \\
\hline
\hline
\noalign{\smallskip}
               &             &        &        &        &                   &          &     \\  
 3300 - 3600   &       300   & 3.50   & 1.07   &   1  & 0.684 $\pm$ 0.049 &  0.096   &   0.00123 $\pm$ 0.00227  \\
 3600 - 3800   &       200   & 1.94   & 1.33   &   1  & 0.606 $\pm$ 0.035 &  0.141   &   0.00083 $\pm$ 0.00158  \\
 3800 - 3937   &       137   & 1.60   & 1.50   &   1  & 0.652 $\pm$ 0.032 &  0.080   &   0.00115 $\pm$ 0.00146  \\
 3946 - 4080   &       134   & 1.19   & 1.11   &   1  & 0.650 $\pm$ 0.021 &  0.124   &  -0.00155 $\pm$ 0.00081  \\
 4080 - 4230   &       150   & 0.91   & 1.34   &   1  & 0.553 $\pm$ 0.021 &  0.182   &   0.00011 $\pm$ 0.00072  \\
 4230 - 4370   &       140   & 0.96   & 1.05   &   1  & 0.590 $\pm$ 0.021 &  0.118   &  -0.00024 $\pm$ 0.00061  \\
 4370 - 4500   &       130   & 0.79   & 1.04   &   0  & 0.544 $\pm$ 0.021 &  0.175   &  -0.00001 $\pm$ 0.00050  \\
 4590 - 4700   &       110   & 0.71   & 1.04   &   0  & 0.492 $\pm$ 0.021 &  0.223   &  -0.00031 $\pm$ 0.00045  \\
 4700 - 4800   &       100   & 0.75   & 1.07   &   0  & 0.466 $\pm$ 0.021 &  0.231   &  -0.00032 $\pm$ 0.00048  \\
 4800 - 4900   &       100   & 0.76   & 1.00   &   0  & 0.394 $\pm$ 0.021 &  0.251   &   0.00031 $\pm$ 0.00046  \\
 4900 - 4980   &        80   & 0.83   & 1.00   &   1  & 0.436 $\pm$ 0.021 &  0.224   &  -0.00012 $\pm$ 0.00051  \\
 4980 - 5060   &        80   & 0.77   & 1.00   &   0  & 0.442 $\pm$ 0.021 &  0.228   &   0.00026 $\pm$ 0.00046  \\
 5060 - 5140   &        80   & 0.88   & 1.00   &   1  & 0.434 $\pm$ 0.021 &  0.228   &  -0.00034 $\pm$ 0.00053  \\
 5140 - 5220   &        80   & 0.89   & 1.11   &   1  & 0.444 $\pm$ 0.021 &  0.206   &  -0.00056 $\pm$ 0.00060  \\
 5220 - 5300   &        80   & 0.81   & 1.06   &   0  & 0.413 $\pm$ 0.021 &  0.221   &   0.00006 $\pm$ 0.00052  \\
 5300 - 5380   &        80   & 0.85   & 1.12   &   1  & 0.430 $\pm$ 0.021 &  0.230   &  -0.00128 $\pm$ 0.00059  \\
 5380 - 5460   &        80   & 0.80   & 1.35   &   1  & 0.384 $\pm$ 0.021 &  0.237   &  -0.00055 $\pm$ 0.00065  \\
 5460 - 5530   &        70   & 0.89   & 1.00   &   0  & 0.407 $\pm$ 0.021 &  0.218   &   0.00024 $\pm$ 0.00054  \\
 5530 - 5600   &        70   & 0.93   & 1.00   &   3  & 0.386 $\pm$ 0.021 &  0.243   &  -0.00179 $\pm$ 0.00111  \\
 5600 - 5720   &       120   & 0.86   & 1.29   &   3  & 0.279 $\pm$ 0.021 &  0.229   &  -0.00170 $\pm$ 0.00143  \\
 5720 - 5800   &        80   & 0.71   & 1.24   &   1  & 0.289 $\pm$ 0.021 &  0.240   &   0.00043 $\pm$ 0.00053  \\
 5800 - 5878   &        78   & 0.71   & 1.21   &   1  & 0.290 $\pm$ 0.021 &  0.239   &   0.00012 $\pm$ 0.00052  \\
 5878 - 5908   &        30   & 1.20   & 1.35   &   1  & 0.257 $\pm$ 0.031 &  0.237   &   0.00077 $\pm$ 0.00097  \\
 5908 - 5990   &        82   & 0.59   & 1.49   &   1  & 0.270 $\pm$ 0.021 &  0.243   &  -0.00041 $\pm$ 0.00054  \\
 5990 - 6070   &        80   & 0.73   & 1.00   &   0  & 0.235 $\pm$ 0.021 &  0.244   &   0.00041 $\pm$ 0.00043  \\
 6070 - 6150   &        80   & 0.65   & 1.00   &   1  & 0.261 $\pm$ 0.021 &  0.234   &   0.00014 $\pm$ 0.00039  \\
 6150 - 6230   &        80   & 0.71   & 1.00   &   1  & 0.263 $\pm$ 0.021 &  0.230   &   0.00008 $\pm$ 0.00042  \\
 6230 - 6310   &        80   & 0.59   & 1.26   &   0  & 0.243 $\pm$ 0.021 &  0.227   &   0.00026 $\pm$ 0.00045  \\
 6310 - 6390   &        80   & 0.57   & 1.00   &   0  & 0.227 $\pm$ 0.019 &  0.236   &  -0.00036 $\pm$ 0.00035  \\
 6390 - 6465   &        75   & 0.58   & 1.05   &   1  & 0.241 $\pm$ 0.018 &  0.233   &  -0.00084 $\pm$ 0.00037  \\
 6465 - 6535   &        70   & 0.60   & 1.10   &   0  & 0.221 $\pm$ 0.021 &  0.236   &  -0.00036 $\pm$ 0.00042  \\
 6535 - 6585   &        50   & 0.85   & 1.24   &   1  & 0.155 $\pm$ 0.022 &  0.264   &   0.00114 $\pm$ 0.00064  \\
 6585 - 6660   &        75   & 0.56   & 1.00   &   1  & 0.206 $\pm$ 0.019 &  0.240   &   0.00006 $\pm$ 0.00034  \\
 6660 - 6740   &        80   & 0.65   & 1.00   &   0  & 0.205 $\pm$ 0.021 &  0.232   &   0.00007 $\pm$ 0.00040  \\
 6740 - 6820   &        80   & 0.66   & 1.10   &   0  & 0.215 $\pm$ 0.023 &  0.228   &  -0.00028 $\pm$ 0.00045  \\
 6820 - 6900   &        80   & 0.62   & 1.12   &   0  & 0.231 $\pm$ 0.022 &  0.229   &  -0.00075 $\pm$ 0.00044  \\
 6900 - 6980   &        80   & 0.72   & 1.48   &   0  & 0.218 $\pm$ 0.029 &  0.232   &  -0.00058 $\pm$ 0.00068  \\
 6980 - 7060   &        80   & 0.68   & 1.00   &   0  & 0.191 $\pm$ 0.021 &  0.224   &  -0.00026 $\pm$ 0.00041  \\
 7060 - 7140   &        80   & 0.72   & 1.01   &   0  & 0.226 $\pm$ 0.022 &  0.227   &  -0.00039 $\pm$ 0.00045  \\
 7140 - 7220   &        80   & 0.64   & 1.60   &   0  & 0.237 $\pm$ 0.028 &  0.221   &  -0.00005 $\pm$ 0.00066  \\
 7220 - 7300   &        80   & 0.79   & 1.49   &   1  & 0.265 $\pm$ 0.020 &  0.228   &  -0.00094 $\pm$ 0.00072  \\
 7300 - 7380   &        80   & 1.17   & 1.95   &   1  & 0.297 $\pm$ 0.037 &  0.219   &  -0.00071 $\pm$ 0.00140  \\
 7380 - 7450   &        70   & 1.65   & 1.87   &   1  & 0.189 $\pm$ 0.051 &  0.223   &   0.00028 $\pm$ 0.00184  \\
 7580 - 7675   &        95   & 1.74   & 1.26   &   0  & 0.291 $\pm$ 0.036 &  0.221   &   0.00048 $\pm$ 0.00137  \\
 7675 - 7710   &        35   & 1.68   & 1.12   &   0  & 0.275 $\pm$ 0.030 &  0.220   &   0.00324 $\pm$ 0.00114  \\
 7710 - 7780   &        70   & 1.56   & 1.00   &   1  & 0.240 $\pm$ 0.025 &  0.225   &   0.00085 $\pm$ 0.00098  \\
 7780 - 7860   &        80   & 1.45   & 1.20   &   0  & 0.267 $\pm$ 0.030 &  0.222   &   0.00141 $\pm$ 0.00119  \\
 7860 - 7950   &        90   & 1.58   & 1.00   &   0  & 0.224 $\pm$ 0.028 &  0.220   &   0.00171 $\pm$ 0.00106  \\
 7950 - 8040   &        90   & 1.21   & 1.00   &   3  & 0.269 $\pm$ 0.020 &  0.224   &   0.00019 $\pm$ 0.00155  \\
 8040 - 8140   &       100   & 1.00   & 1.34   &   0  & 0.206 $\pm$ 0.023 &  0.222   &  -0.00112 $\pm$ 0.00089  \\
 8140 - 8240   &       100   & 1.29   & 1.11   &   3  & 0.143 $\pm$ 0.026 &  0.217   &  -0.00150 $\pm$ 0.00188  \\
 8240 - 8350   &       110   & 1.02   & 1.00   &   1  & 0.198 $\pm$ 0.027 &  0.220   &   0.00072 $\pm$ 0.00062  \\
 8350 - 8460   &       110   & 1.14   & 1.14   &   0  & 0.177 $\pm$ 0.022 &  0.217   &   0.00046 $\pm$ 0.00079  \\
 8460 - 8560   &       100   & 1.17   & 1.00   &   0  & 0.156 $\pm$ 0.026 &  0.210   &  -0.00006 $\pm$ 0.00074  \\
 8560 - 8640   &        80   & 1.06   & 1.21   &   3  & 0.179 $\pm$ 0.021 &  0.217   &  -0.00023 $\pm$ 0.00170  \\
 8640 - 8730   &        90   & 1.21   & 1.00   &   0  & 0.171 $\pm$ 0.021 &  0.215   &  -0.00147 $\pm$ 0.00077  \\
 8730 - 8850   &       120   & 1.21   & 1.00   &   1  & 0.145 $\pm$ 0.024 &  0.219   &   0.00136 $\pm$ 0.00076  \\
 8850 - 9000   &       150   & 1.13   & 1.28   &   0  & 0.184 $\pm$ 0.024 &  0.219   &   0.00066 $\pm$ 0.00092  \\
 9000 - 9150   &       150   & 1.12   & 1.00   &   0  & 0.167 $\pm$ 0.029 &  0.214   &  -0.00031 $\pm$ 0.00070  \\
 9150 - 9350   &       200   & 1.05   & 1.00   &   0  & 0.172 $\pm$ 0.028 &  0.217   &   0.00188 $\pm$ 0.00065  \\
 9350 - 9600   &       250   & 1.60   & 1.14   &   0  & 0.154 $\pm$ 0.032 &  0.216   &  -0.00001 $\pm$ 0.00118  \\
 9600 - 10000  &       400   & 2.13   & 1.00   &   1  & 0.203 $\pm$ 0.036 &  0.209   &  -0.00166 $\pm$ 0.00136  \\
\hline                                                                         
\end{tabular}                                                                  
\end{center}
\end{table*}
}

\section{Discussion}

\subsection{Lack of pressure-broadened alkali absorption}

The transmission spectrum of HAT-P-32b derived in this work shows no significant atomic or molecular absorption features. The lack of features predicted by cloud-free solar-composition models \citep{Fortney2010} of Na, K, or TiO confirms the result obtained by G13 and is best explained by a layer of clouds or haze obscuring the spectral signatures. Within the current sample of hot Jupiters observed by transmission spectroscopy at optical wavelengths, the vast majority show no or weakened alkali features compared to cloud-free solar-composition models. Recently, \cite{Fischer2016} published the result for WASP-39b, which is an exception in that it actually shows the pressure-broadened alkali wings indicative of a clear atmosphere. \cite{Sing2016} gave evidence that the lack of observed features is linked to clouds/hazes rather than being caused by subsolar abundances. However, the formation of these particles in the atmosphere is not yet fully understood and no clear correlation of their occurrence with atmospheric temperature, planetary surface gravity or stellar insolation emerges from the sample of existing observations. \cite{Fortney2005} predicted the condensate opacity to be important for the observational technique of transmission spectroscopy because of the slant viewing geometry and the resulting long path of the star light through the planetary atmosphere. Trace species of condensates can make the atmosphere optically thick and hamper the detection of further atomic or molecular absorption features.

\subsection{Lack of TiO}
Indications of a temperature inversion at the permanent dayside were found for HAT-P-32b by \cite{Zhao2014}, possibly caused by TiO. However, the cloud-free solar-composition model of 1750~K by \cite{Fortney2010}, which is dominated by TiO absorption, is ruled out to high significance at the planetary terminator probed by our measurements. The large discrepancy between measurements and model is mainly caused by the lack of the predicted blueward drop in absorption (see Figure \ref{plot_transspec_2pan}). However, the blueward drop might be damped by the opacity of clouds or hazes. Several absorption features of TiO at redder wavelengths might be measurable instead, in spite of clouds/hazes, because the atmosphere becomes opaque at very high altitudes as a result of TiO absorption \citep{Fortney2010}. We searched for the bandheads at $\sim$\,7200~\AA{}, $\sim$\,7700~\AA{}, and $\sim$\,8500~\AA{}, and can significantly rule out additional absorption. The lack of TiO could be explained by a rainout of TiO bound in condensates at lower altitudes or condensation on the colder night side \citep{Spiegel2009} and would be in agreement with the lack of TiO absorption in transmission spectra of other hot Jupiters. 

\subsection{Tentative indication of potassium}
In both the original transmission spectrum and the spectrum after chromatic mode correction we detect an indication of additional absorption by potassium. In the original spectrum the additional absorption is $\Delta\mathrm{k}\,=\,0.0032\,\pm\,0.0011$ compared to the adjacent channels, and $\Delta\mathrm{k}\,=\,0.0025\,\pm\,0.0010$ compared to the best-fit linear function over wavelength, which corresponds to a significance of 2.9 and 2.5\,$\sigma$, respectively. In the chromatic mode corrected spectrum the additional absorption is 0.0032\,$\pm$\,0.0011 or 2.8\,$\sigma$ compared to the four channels blueward and four channels redward. Our analysis concentrates on the line core at 7699~\AA{} of the potassium doublet because the photometry of the other line core at 7665~\AA{} will be influenced by the telluric Fraunhofer A band. We tested several channel widths around the 7699~\AA{} line from 10 to 60~\AA{} in 10~\AA{} steps and found no increase in absorption toward narrower channels, which would be expected 
since wider channels dilute the signal of the line core. We also tested different channel widths around the Na doublet at $\sim$\,5893~\AA{} and H$\alpha$ at 6563~\AA{} and found consistent values without an increase in absorption for narrower channels. 

An interpretation for finding K but no Na in the atmosphere of HAT-P-32b would not be trivial. Photo-ionization favors a depletion of K I over Na I \citep{Sing2015}. Therefore, it would not be a plausible explanation for a low Na/K ratio. Both elements should not condense out at the equilibrium temperature of $\sim$\,1800~K since Na forms condensates at $\sim$\,1000~K, while K condenses around 800~K \citep{Lodders2003}. 

More observational data are needed to confirm an excess absorption in K before a conclusive statement can be drawn. 

\subsection{Spectral slope caused by scattering?}

The derived spectrum shows an increase in opacity toward bluer wavelengths covering the entire observed spectral range from 3300 to 10000~\AA{}. A fit of a flat line gives a $\chi^2$ of 111.2 (DOF=61), which corresponds to a one-tailed probability of $1 \times 10^{-4}$. We verified that this slope is not caused by the treatment of the limb-darkening parameters by two additional transit fit runs. We included the two LDCs as fixed to their theoretical values; in another fit we left both LDC free. In both cases, the slope was not significantly different from the transmission spectrum of Figure \ref{plot_transspec_2pan} (Table \ref{tab_62lcs}), for which the linear LDC was fitted and the quadratic LDC was fixed. It also made no significant difference whether we fixed $i$ and $R_{\star}$ to the values used by G13 or the red or blue light curve values of Table \ref{transitparam_H32}.

G13 does not discuss the significance of a slope in their transmission spectrum. Using the G13 planetary spectrum to fit a linear function over wavelength in \AA{}, we find a slope of $(-3.95\,\pm\,1.10)\,\times\,10^{-7}$, which is in good agreement with the derived slope of $(-3.77\,\pm\,0.89)\,\times\,10^{-7}$ from the transmission spectrum of this work. However, the significance of the slope is only marginally above 3\,$\sigma$, and we are cautious about announcing the detection of a spectral gradient. Follow-up observations are needed in order to  rule out undetected red noise as its source. The follow-up can potentially be done by multiple transit observations with small to medium class telescopes using broad-band filters \citep[e.g.,][]{MallonnH12}, and we collected more than a dozen each at blue and red optical filters. Their analysis will be presented in a subsequent paper. Here, we present viable interpretations of the measured slope if it can be 
confirmed by future observations.

\subsubsection{Permanent spot filling factor}

In Section \ref{chap_res_lotephot} we presented the results of our two-band monitoring campaign and showed that there are no photometric variations, for example,  caused by  spots that might significantly influence our transmission spectrum. However, the monitoring data  cannot rule out permanently visible spots either in a latitudinal band or as polar spots, which could in principle produce a blueward slope in the spectrum as observed here \citep{McCullough2014}. Such permanently visible spots cause no photometric variation and would be invisible in our monitoring light curves. Similarly to the estimation of the spot correction in Section \ref{chap_res_lotephot}, we estimated what spot filling factor would be needed to explain the observed slope in the transmission spectrum purely with unocculted spots. We assumed a spot temperature 2000~K lower than the photosphere according to the empirical trend of larger spot-photosphere temperature contrast for hotter stars listed in \cite{Berdyugina2005}. For 
simplicity, we assumed black body radiation for spot and photosphere and found that a spot filling factor of about 5\,\% would be needed to cause the observed slope. However, it appears unlikely that such a large spot fraction is distributed homogeneously enough in a belt to cause a periodic photometric variation below our upper limit of 1\,mmag derived in Section \ref{chap_res_lotephot}. Furthermore, such a filling factor would be unusually large regarding the shallow convection zone of F-type main-sequence  stars, the advanced age of HAT-P-32, and the lack of significant chromospheric emission in its Ca II H$\&$K line cores \citep{Hartman2011}. It is also not expected  according to the statistics of observed variability in F-type main-sequence stars derived from the Kepler satellite \citep{McQuillan2012}. Therefore, we conclude that the slope in the derived transmission spectrum is not caused by activity of the host star.

\subsubsection{Additional third-light contribution}

We derived values for the third-light contribution of the visible M dwarf companion in Section \ref{Chap_H32_Mdwarf} and corrected the transmission spectrum for its influence. The observed, tentative, gradient in the spectrum could be brought into rough agreement with a flat line by a second similar correction of about 150\,\% amplitude of the first. Therefore, we ask whether the measured slope could be caused by an additional stellar companion. In fact, \cite{Knutson2014} detected a trend in the RV data and interpreted it as  an additional companion of either planetary or stellar size. The K-band high-resolution spectroscopy of HAT-P-32 of \cite{Piskorz2015} ruled out an object hotter than 3500\,K. Thus, the object causing the RV trend cannot be luminous enough to be responsible for the observed slope in the transmission spectrum. An AO search by \cite{Ngo2015} did not detect an additional unknown stellar object either gravitationally bound or by chance aligned along the line of sight. We conclude that the 
slope of the derived transmission spectrum is most likely not caused by third light from an additional stellar object.

We conclude that there is no additional stellar object along the line of sight that is luminous enough to cause a spectral slope in the transmission spectrum comparable in amplitude to the one tentatively measured.

\subsubsection{Scattering processes in the planetary atmosphere}
\label{Sec_Disc_scatter}
Following the conclusions of the previous sections, we assume the measured slope in the transmission spectrum to be a scattering feature intrinsic to the planetary atmosphere. It extends over $\sim$\,2~scale heights of $\sim$\,1000~km, which is shallower than the Rayleigh slope of HD\,189733b that extends over $\sim$\,4~scale heights from 3500 to 10000~\AA{} \citep{Sing2011}. The extinction cross section $\sigma$ of scattering and absorption processes of an atmospheric opacity source was described by \cite{LecavelierDesEtangs2008} with a  power law of index $\alpha$, i.e., $\sigma\,=\,\sigma(\lambda / \lambda_0)^{\alpha}$. The slope of the planetary transmission spectrum is then proportional to the product $\alpha T$,
\begin{equation}
\alpha T\,=\, \frac{\mu g}{k}\,\frac{\mathrm{d}(R_p/R_{\star})}{\mathrm{d}\,\mathrm{ln} \lambda} .
\label{equ_alphaT}
\end{equation}
A fit to the HAT-P-32b spectrum results in $\alpha T = -3560 \pm 910$~K with a $\chi^2$ of 82.8 with~60 DOF. If we assume that the slope in the transmission spectrum is caused by Rayleigh scattering ($\alpha\,=\,-4$), the derived temperature would be 890\,$\pm$\,228~K. Transmission spectroscopy probes the planetary terminator, i.e., the temperature derived here is a value mid-way between the dayside and the nightside temperature. The measured brightness temperature of the dayside is $\sim$\,2050~K \citep{Zhao2014}, the planet's equilibrium temperature $\sim$\,1800~K \citep{Hartman2011}. Although \cite{Zhao2014} found indications of a rather weak energy redistribution from the dayside to the nightside, the derived terminator temperature seems too low because it would indicate a nightside temperature of only a few hundred K, far below any measured nightside temperatures of hot Jupiters so far. A similar situation of a probably underestimated temperature derived from the slope assuming Rayleigh scattering was 
discussed for WASP-12b by \cite{Sing2013}, who found aerosols causing Mie scattering to be a valuable alternative interpretation. Such aerosols can be formed by condensation or by photochemical processes. Condensate materials for the temperature regime of HAT-P-32b could be corundum (Al$_2$O$_3$) or iron oxide (Fe$_2$O$_3$) \citep{Lodders2003}. A fit of Mie scattering dust models suffers from a degeneracy between the grain size and the atmospheric temperature \citep{Sing2013}. Therefore, we do not attempt a retrieval of the atmospheric parameters.

A source of scattering in the planetary atmosphere could, in principle, also be hydrogen molecules in gas phase. However, this explanation would need a cloud-free and haze-free atmosphere for which we expect absorption signatures of Na, K, and H$_2$O in our spectral range. A depletion of all three species at the wide range of pressures probed by the measurement in the case of a clear atmosphere is unlikely.

Another valuable interpretation of the spectral slope including Rayleigh scattering from aerosols was mentioned by \cite{Sing2013} for WASP-12b, drawn from insights of the solar system. It is possible that the scale height of aerosols is smaller than the scale height of the atmospheric gas, a property discovered for ammonia clouds in the Jovian atmosphere \citep{Brooke1998}. This scale height ratio would result in a smaller Rayleigh slope without the necessity of a temperature much cooler than the equilibrium temperature.

\section{Conclusions}

We observed one transit event of the inflated hot Jupiter HAT-P-32b with MODS at the LBT and measured the planetary transmission spectrum. MODS disperses the light toward a blue and a red arm with individual CCD detectors and offers a wavelength coverage from 3300~\AA{} to 10000~\AA{} in a single exposure. We divided each target spectrum into 62 wavelength channels, 19 channels in the blue region and 43 in the red. For each channel we created differential light curves. All light curves redder than 5600~\AA{} were corrected for the third-light contribution of a nearby M dwarf. We also took the potential influence of photospheric
brightness inhomogeneities of the host star into account. The averaged blue and red light curves were able to reproduce the previously known transit parameters with a model fit to within 1\,$\sigma$ deviation.

The transmission spectrum derived from the set of 62 light curves over wavelength lacks pressure-broadened absorption features of sodium and potassium and shows no spectral features of TiO  at blue or at red wavelengths. In agreement with previous transmission spectroscopy work of HAT-P-32b by G13, the measurements are interpreted as an atmosphere that is not free of clouds/hazes and which lacks TiO in gas phase. Furthermore, the spectrum of this work shows a tentative slope of increasing effective planet size from the red toward the blue, which we argue is not likely  caused by the host star or other stars in the aperture. If confirmed by follow-up observations, it is likely caused by scattering processes in the planetary atmosphere. The amplitude of the slope is weaker than predicted by Rayleigh scattering when the scale height is estimated using the equilibrium temperature of the planet. We find either Mie scattering caused by aerosols to be a plausible explanation or Rayleigh scattering with an aerosol - 
gas scale height ratio significantly lower than unity.
 
We measure an excess absorption in the line core of potassium at 2.8\,$\sigma$ confidence, but argue that additional observations are needed to confirm this result. No excess absorption was found in the line cores of sodium and H$\alpha$, perhaps because the features are diluted by clouds or hazes. The transmission spectrum of HAT-P-32b from the near-UV to the near-IR confirms the importance of clouds/hazes in the interpretation of hot Jupiter transmission spectra. A study by our team to verify the tentative scattering feature in the planetary atmosphere by broad-band spectrophotometry is  in progress.

\begin{acknowledgements}
We thank Antonio Claret for providing the theoretical limb-darkening coefficients and Jonathan Fortney for providing the planetary atmospheric models. We thank Barry Rothberg for performing the observations at the LBT, and Olga Kuhn and Dave Thompson for the helpful discussions regarding their preparation. We are grateful to Thomas Granzer for all help reagarding the STELLA telescope. We thank Carolina von Essen, Elyar Sedaghati, J\"org Weingrill, and Andreas K\"unstler for discussions on the manuscript. This research has made use of the SIMBAD database and the VizieR catalog access tool, operated at CDS, Strasbourg, France.
\end{acknowledgements}

%
\bibliographystyle{aa} 
\bibliography{myfile} 
%

\end{document}